\documentclass[
    ,final            
  ]
  {aipproc}

\layoutstyle{8x11double}

\begin{document}

\title{The Full Spectrum Galactic Terrarium: MHz to TeV Observations of Various Critters}

\classification{95.85.Pw, 95.85.Nv, 95.85.Bh, 97.60.Gb, 97.60.Jd, 97.80.Jp, 98.38.Dq, 98.38.Mz}
\keywords      {pulsar, pulsar wind nebula, supernova remnant, EGRET, FERMI, GLAST, gamma-ray, VLA, XMM-Newton, Binary, ATCA, PSR J2021+3651, HESS, MILAGRO, MGRO J1908+06, RCW 49, LS 5039}

\author{Mallory S.E. Roberts}{
  address={Eureka Scientific, Inc., 2452 Delmer Street, Suite 100, Oakland, CA 94602-3017 },
  email={malloryr@gmail.com}
}

\author{C. Brogan}{
  address={National Radio Astronomy Observatory, 520 Edgemont Road, Charlottesville, VA 22903},
}

\author{S. Ransom}{
  address={National Radio Astronomy Observatory, 520 Edgemont Road, Charlottesville, VA 22903},
}

\author{M. Lyutikov}{
  address={Department of Physics, Purdue University, West Lafayette, IN 47907, USA}
}

\author{E. de O\~na Wilhelmi}{
  address={Astroparticule et Cosmologie (APC), CNRS, Universite Paris 7 Denis Diderot, 10, Paris, France}
}

\author{A. Djannati-Ata\"i}{
  address={Astroparticule et Cosmologie (APC), CNRS, Universite Paris 7 Denis Diderot, 10, Paris, France}
}

\author{R. Terrier}{
  address={Astroparticule et Cosmologie (APC), CNRS, Universite Paris 7 Denis Diderot, 10, Paris, France}
}

\author{S.M. Dougherty}{
  address={National Research Council of Canada, Herzberg Institute for Astrophysics, Dominion Radio Astrophysical Observatory, PO Box 248, Penticton, British Columbia V2A 6J9, Canada}
}

\author{E.D. Grundstrom}{
  address={Physics and Astronomy Department, Vanderbilt University, 6301 Stevenson Center, Nashville, TN 37235}
}

\author{J.W.T. Hessels}{
  address={Astronomical Institute ``Anton Pannekoek,'' University of Amsterdam, 1098 SJ Amsterdam, Netherlands}
}

\author{S. Johnston}{
  address={Australia Telescope National Facility, CSIRO, PO Box 76, Epping, NSW 1710, Australia}
}

\author{M.V. McSwain}{
  address={Department of Physics, Lehigh University, Bethlehem, PA 18015}
}

\author{P.S. Ray}{
  address={Space Science Division, Naval Research Laboratory, Washington, DC 20375-5352}
}

\author{K.S. Wood}{
  address={Space Science Division, Naval Research Laboratory, Washington, DC 20375-5352}
}

\author{G.G. Pooley}{
  address={Cavendish Laboratory, Cambridge University}
}

\author{A. Weinstein}{
  address={Department of Physics and Astronomy, University of California, Los Angeles, CA 90095}
}

\begin{abstract}
Multi-wavelength studies at radio, infrared, 
optical, X-ray, and TeV wavelengths have discovered probable counterparts to many Galactic sources of GeV emission detected by EGRET. These include 
pulsar wind nebulae, high mass X-ray binaries, and mixed morphology supernova remnants.
Here we provide an overview of the observational properties of Galactic sources which emit across 19 orders of magnitude in energy. We also present new observations of several sources.
\end{abstract}

\maketitle


\section{All Across the Electromagnetic Spectrum}

For many years, the Crab nebula was the only Galactic source detected across the entire spectrum, from radio to TeV energies \citep{wcf+89}.
The EGRET telescope on board the Compton Gamma-Ray Observatory detected dozens of Galactic sources at GeV energies \citep{lm97}, most of which
were not firmly identified by the end of the mission. Since then, multi-wavelength studies of the EGRET error boxes at radio 
optical and X-ray wavelengths have discovered probable counterparts for most of the sources bright above 1 GeV, most of them the products of supernovae (pulsars,  
pulsar wind nebulae (PWN), and supernova remnants (SNR), \citep{r08}).  Furthermore, surveys of the Galactic plane by HESS, MILAGRO, and VERITAS (among others) have discovered TeV emitting regions
which can plausibly be associated with many of these sources.  Here we provide an overview of the observational properties of sources which emit across 19 
orders of magnitude in energy.

\section{Young Pulsars and Pulsar Wind Nebulae}

Young spin-powered pulsars were the only firmly identified (through their pulsations) class of Galactic GeV sources during the EGRET era. Pulsed emission from the longest
radio wavelengths up to a few tens of GeV has been detected, although in general the $\gamma-$ray cut-off for pulsed emission seems to be in the few GeV range \citep[see][for a general review of pulsar and pulsar wind nebulae properties]{krh06}.  
The  winds of young pulsars create nebulae which emit synchrotron radiation from radio through soft gamma-rays, 
and through inverse Compton scattering produce photons with energies up to tens of TeV.  
The transition from synchrotron to inverse Compton emission in PWN seems to be in the $\sim 0.01-1$ GeV range, although only for the Crab and Vela is the synchrotron 
cut-off energy fairly well established. The cut-off energies of both the pulsed and PWN emission should be greatly constrained by the Fermi (formerly known as GLAST) 
LAT over the coming years. 

Inverse Compton (IC) emission is a very efficient means of producing high-energy photons from an accelerated particle with a large fraction of the electron energy transferred
to the seed photon in a single scatter. The peak of the IC emission is typically 5-10 orders of magnitude higher in energy than the corresponding synchrotron 
radiation. It is also not dependent on magnetic field strength, and so can produce significant amounts of radiation even in regions with very low magnetic field strength. 
The implication of this for emission from PWN is that the synchrotron lifetimes of the IC emitting electrons can be very long, and it is not necessarily the case 
that where the synchrotron emission is bright, the IC emission is bright. In fact, it is typically the case that the TeV emission is significantly offset from the current 
position of the pulsar. This offset is caused by either the motion of the pulsar or the interaction of the supernova remnant reverse shock with the PWN. 

The MILAGRO telescope detected two previously unknown regions of significant emission above 10 TeV \citep{aab+07}: a complex of sources in the 
Cygnus region with the brightest peak known as MGRO J2019+37 and another extended source called MGRO J1908+06. The nearest EGRET source to MGRO J2019+37
is GeV J2020+3658 which is associated with the young pulsar and PWN PSR J2021+3651 \citep{hrr+04}, which is near the central region of the Cygnus cross. For this reason, 
and the fact that the pulsar was discovered with Arecibo in Puerto Rico, we sometimes refer to it as Cisne (swan in Spanish).  

We have performed X-ray 
observations with XMM-Newton and 20cm radio observations with the VLA of the region around PSR J2021+3651. Outside of the bright inner X-ray nebula there structured hard, faint X-ray emission extending out $\sim 10-15$ arcminutes. The brightest region
of this larger X-ray nebula is an extension to the west of about $8^{\prime}$ in length. The VLA image shows a radio nebula coincident with this X-ray extension, with a 
suggestion of a conical morphology extending out around $10^{\prime}$, at which point there is a decrease in surface brightness and a further broadenig of the nebula. 
The radio nebula extends out at least $20^{\prime}$ to the west, right to the center of the best fit ellipse to MGRO J2019+37. This connection, for any reasonable 
distance to PSR J2021+3651 ($> 6$ kpc, see discussion in \citep{hrr+04}) results in a TeV luminosity several times that of the Crab nebula, 
making this the most luminous TeV source in the Galaxy.  If we include the EGRET flux, the broad-band spectral energy distribution of the PWN would appear to 
peak in the GeV range. However, much if not most of the emission seen by EGRET is undoubtably magnetospheric pulsed emission, therefore it is likely the synchrotron 
emitting PWN cuts off in the MeV range. 
Unfortunately, VERITAS has not yet been able to confirm the TeV source, but should ultimately give a much more detailed understanding of the TeV emission and 
its morphological connection to the PWN of PSR J2021+3651.

\begin{figure}
  \includegraphics[height=.25\textheight]{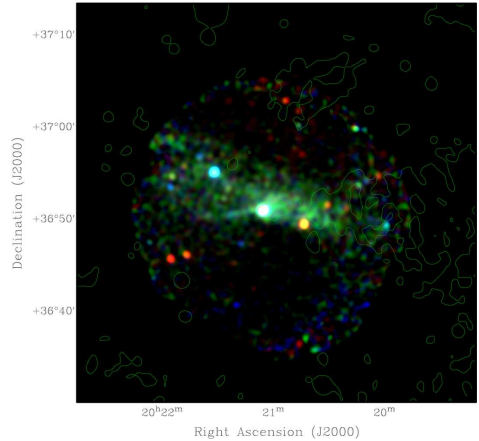}
  \caption{XMM-Newton X-ray image of Cisne, the PWN around PSR J2021+3651 (for color version, red=0.5-1.5 keV, green=1.5-2.5, blue=2.5-7.5) with VLA contours}
\end{figure}

\begin{figure}
  \caption{VLA 20cm image of Cisne PWN with XMM-Newton X-ray contours}
  \includegraphics[height=.25\textheight]{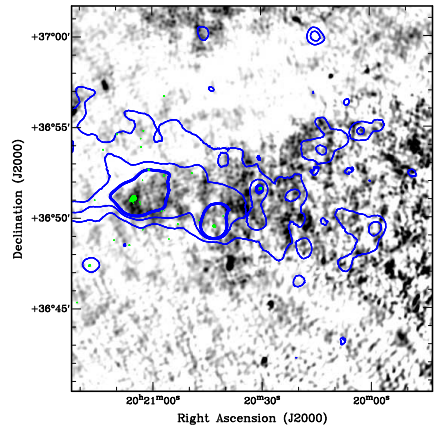}
\end{figure}

\begin{figure}
  \includegraphics[height=.3\textheight, angle=270]{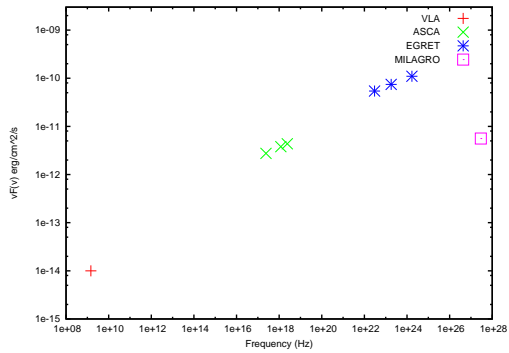}
  \caption{Spectral Energy Distribution of Cisne. Note that the EGRET flux undoubtably has a significant pulsed component.}
\end{figure}

MGRO J1908+06 is near GeV J1907+0557, for which a potentially extended X-ray counterpart was discovered with ASCA \citep{rrk01}. A brief (10~ks) Chandra ACIS observation 
showed most of the flux is from a hard (photon index $\sim 1$) moderately absorbed (nH $\sim 2 \times 10^{22} {\rm cm}^{-2}$) point source with no compact nebula and just a faint hint of the larger nebula seen by ASCA. A moderately 
deep I band image with the MDM 2.4m telescope shows no evidence of an optical counterpart.
Both HESS and VERITAS (see other contributions, this proceedings) have verified and resolved the MILAGRO source, with the ASCA source at the southern edge. AGILE data of 
the region verifies the GeV emission is consistent with the position of GeV J1907+0557 and suggests the third EGRET catalog source 3EG J1903+0550 was a 
composite of the GeV source and another, unrelated source. 

\begin{figure}
  \includegraphics[height=.25\textheight, angle=270]{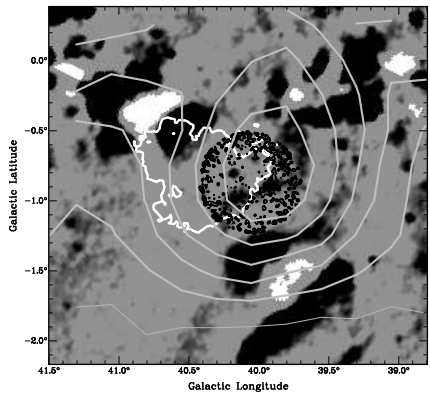}
  \caption{GB6 4850MHz image \citep{cbs+94} of MGRO J1908+06 region. The white contour is the HESS source, the
black contour is the ASCA 2-10 keV image (nb. that the circle is due to exposure correction
of noise at the edge of the FOV, the source under discussion lies along the outside edge of the
HESS contour) and the grey contour is the AGILE $> 300$ MeV image.}
\end{figure}

Several of the unidentified EGRET sources
containing PWN appear to be variable on
timescales of months according to the
Nolan et al. (2003) variability test \citep{ntgm03}
with $\tau = F_{rms}/F_{mean}\sim 1$. This suggests the
GeV flux may be partially due to particle acceleration
in the inner nebula rather than being pulsed emisssion from the magnetosphere. 
Some of these sources (eg. the Rabbit/GeV J1417-6100 and the Eel/GeV J1825-1310 nebulae, see \citep{r07})  appear to emit TeV emission as well.
However, three potentially variable GeV sources, CTA 1, GeV J1809-2327 (Taz) and SNR W44, have PWN inside of mixed-morphoogy SNR \citep{rb08}. 
These supernova remnants have radio shells but are filled with 
thermal X-rays. Most SNR of this type also contain young pulsars with pulsar wind 
nebulae. Two mixed-morphology SNR associated with GeV sources, W28 and IC 443, have had 
TeV hotspots reported \citep{aab+08,aaa+07}, suggesting the three containing variable GeV sources are likely to be TeV emitters as well. 

Complicating our understanding of these sources are their environments. Many are near molecular clouds, which may both enhance the
the production of $\gamma-$ray emission by particles accelerated in pulsar winds and SNR shocks as well as produce $\gamma-$ray emission 
on their own. In addition, there have been a few claims of extended TeV sources associated with massive stars or colliding wind binaries contained
within molecular clouds despite the TeV emission being somewhat offset from the stellar systems. One should keep in mind that 
the bright diffuse radio emission from molecular clouds can obscure PWN or SNR shells. We have performed ATCA radio 
imaging of one such source, the Westerlund 2 complex (also known as RCW 49) near GeV J1025$-$5809 which is coincident with HESS J1023$-$575 \citep{aab+07b}. While much of the 
radio emission can be associated with mid-infrared emission seen by MSX, there are some arcs in the ATCA image that do not correlate with infrared structures. This suggests a non-thermal, i.e. SNR shell, identification for these arcs. 

\begin{figure}
  \includegraphics[height=.4\textheight]{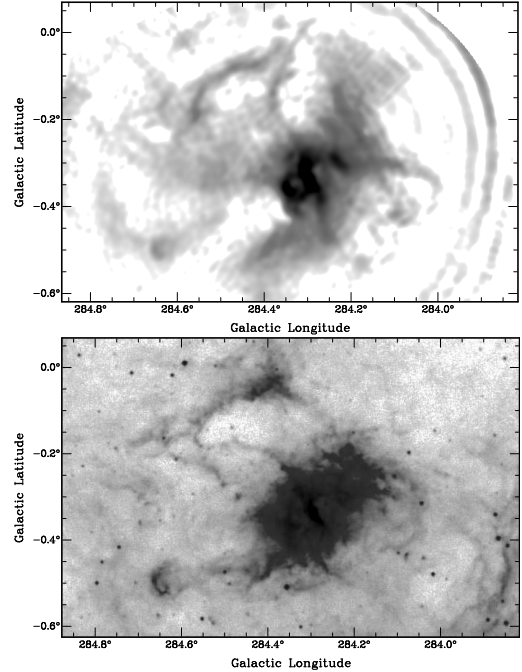}
  \caption{Top: ATCA 20cm image of RCW 49. Bottom: MSX 8.3$\mu$ image.}
\end{figure}

\section{Radio-Loud High Mass X-Ray Binaries}

There are at least 3 strange radio emitting binary systems with massive stars in eccentric orbits. 
PSR 1259-63 is a young pulsar in a 3.6 year highly eccentric 
orbit around a Be Star. Near periastron, it becomes a 
moderately bright unpulsed broad-band (radio-TeV) 
source (although not yet seen in GeV) \citep{jbwm05,nc07}. LSI+61 303 
and LS 5039 are in much smaller eccentric orbits (26.5 and 
3.9 days \citep{gcg+07,crp+05}) which have moderately bright hard X-ray emission but with luminosities about 2-3 orders of magnitude smaller than
known accreting systems.  They also have moderately bright and variable radio emission (on the order of 100 mJy) extended on milliarcsecond scales and are coincident with point sources of TeV emission
\citep[see][for an overview]{dub06}. 
Their emission appears modulated at the orbital period at all frequencies. 
Pulsations from LS 5039 and  LSI+61 303 have not yet been detected. 
We are performing deep, high frequency pulse searches 
with the GBT. Since both of these sources are prime targets for the Fermi telescope, we are monitoring these two sources twice weekly in X-rays with RXTE.  In addition, there are supporting optical and radio monitoring observations planned for LSI+61 303 and LS 5039 lasting through the first year of GLAST. 

\section{A Catalog of Galactic Sources}

The space here is too limited to discuss all of the multiwavelength data of the bright Galactic GeV sources that EGRET detected. An online
catalog of multiwavelength observations of Galactic EGRET sources is being produced by M.S.E. Roberts, and can be found at: 
\url{http://www.physics.mcgill.ca/\~pulsar/unidcat.html} 
This catalog is an expansion of the ASCA Catalog of Potential Counterparts of GeV source \citep{rrk01} and is a work in progress.




\begin{theacknowledgments}
This research has made use of data obtained from the High Energy Astrophysics Science Archive Research Center (HEASARC), provided by NASA's Goddard Space Flight Center.
The National Radio Astronomy Observatory Very Large Array is a facility of the National Science Foundation operated under cooperative agreement by Associated Universities, Inc.
This research made use of data products from the Midcourse Space
Experiment,  processing of which data was funded by the Ballistic
Missile Defense Organization with additional support from NASA
Office of Space Science. The Australia Telescope Compact Array is part of the 
Australia Telescope which is funded by the Commonwealth of Australia for operation as a 
National Facility managed by CSIRO. Based on observations obtained with XMM-Newton, an ESA science mission with instruments and contributions directly funded by ESA Member States and NASA.
This work was partially supported by SAO Grant No. G07-8072 and GO6-7136X, NASA Grant No. NNX08AV70G and NNG06EJ54P. G. McSwain acknowledges institutional support from Lehigh University. M. Roberts 
would like to thank the CNRS and the APC for support during his visit. 
\end{theacknowledgments}



\bibliographystyle{aipproc}   

\bibliography{myrefs,journals_apj}

\begin{thebibliography}{18}
\expandafter\ifx\csname natexlab\endcsname\relax\def\natexlab#1{#1}\fi
\providecommand{\enquote}[1]{``#1''}
\expandafter\ifx\csname url\endcsname\relax
  \def\url#1{\texttt{#1}}\fi
\expandafter\ifx\csname urlprefix\endcsname\relax\def\urlprefix{URL }\fi
\providecommand{\eprint}[2][]{\url{#2}}

\bibitem[{Weekes} et~al.(1989)]{wcf+89}
T.~C. {Weekes}, et al.
\emph{\apj} \textbf{342}, 379--395 (1989).

\bibitem[{Lamb} and {Macomb}(1997)]{lm97}
R.~C. {Lamb}, and D.~J. {Macomb}, \emph{\apj} \textbf{488}, 872--+ (1997).

\bibitem[{Roberts}(2008)]{r08}
M.~S.~E. {Roberts}, \enquote{{Pulsars Everywhere! A Galactic EGRET Source
  Retrospective},} in \emph{40 Years of Pulsars: Millisecond Pulsars, Magnetars
  and More}, edited by C.~{Bassa}, Z.~{Wang}, A.~{Cumming}, and V.~M. {Kaspi},
  2008, vol. 983 of \emph{American Institute of Physics Conference Series}, pp.
  621--623.

\bibitem[{Kaspi} et~al.(2006)]{krh06}
V.~M. {Kaspi}, M.~S.~E. {Roberts}, and A.~K. {Harding}, \emph{{Isolated neutron
  stars}}, Compact stellar X-ray sources, 2006, pp. 279--339.

\bibitem[{Abdo} et~al.(2007)]{aab+07}
A.~A. {Abdo}, et al.
\emph{\apjl} \textbf{664}, L91--L94 (2007)

\bibitem[{Hessels} et~al.(2004)]{hrr+04}
J.~W.~T. {Hessels}, M.~S.~E. {Roberts}, S.~M. {Ransom}, V.~M. {Kaspi}, R.~W.
  {Romani}, C.~. {Ng}, P.~C.~C. {Freire}, and B.~M. {Gaensler}, \emph{\apj}
  \textbf{612}, 389--397 (2004).

\bibitem[{Roberts} et~al.(2001)]{rrk01}
M.~S.~E. {Roberts}, R.~W. {Romani}, and N.~{Kawai}, \emph{\apjs} \textbf{133},
  451--465 (2001)

\bibitem[{Condon} et~al.(1994)]{cbs+94}
J.~J. {Condon}, J.~J. {Broderick}, G.~A. {Seielstad}, K.~{Douglas}, and P.~C.
  {Gregory}, \emph{\aj} \textbf{107}, 1829--1833 (1994).

\bibitem[{Nolan} et~al.(2003)]{ntgm03}
P.~L. {Nolan}, W.~F. {Tompkins}, I.~A. {Grenier}, and P.~F. {Michelson},
  \emph{\apj} \textbf{597}, 615--627 (2003).

\bibitem[{Roberts}(2007)]{r07}
M.~S.~E. {Roberts}, \enquote{{What Will Be The Brightest GLAST Sources in the
  Galaxy?},} in \emph{The First GLAST Symposium}, edited by S.~{Ritz},
  P.~{Michelson}, and C.~A. {Meegan}, 2007, vol. 921 of \emph{American
  Institute of Physics Conference Series}, pp. 385--386.

\bibitem[Roberts \& Brogan(2008)]{rb08} Roberts, M.~S.~E., \& Brogan, C.~L.\ 2008, \apj, 681, 320 

\bibitem[{Aharonian} et~al.(2008)]{aab+08}
F.~{Aharonian}, et al.
\emph{\aap} \textbf{481}, 401--410 (2008).

\bibitem[{Albert} et~al.(2007)]{aaa+07}
J.~{Albert}, et al.
\emph{\apjl} \textbf{664}, L87--L90 (2007)

\bibitem[{Aharonian} et~al.(2007)]{aab+07b}
F.~{Aharonian}, et al.
\emph{\aap} \textbf{467}, 1075--1080 (2007)

\bibitem[{Johnston} et~al.(2005)]{jbwm05}
S.~{Johnston}, L.~{Ball}, N.~{Wang}, and R.~N. {Manchester}, \emph{\mnras}
  \textbf{358}, 1069--1075 (2005)

\bibitem[{Neronov} and {Chernyakova}(2007)]{nc07}
A.~{Neronov}, and M.~{Chernyakova}, \emph{\apss} \textbf{309}, 253--259 (2007)

\bibitem[{Grundstrom} et~al.(2007)]{gcg+07}
E.~D. {Grundstrom}, S.~M. {Caballero-Nieves}, D.~R. {Gies}, W.~{Huang}, M.~V.
  {McSwain}, S.~E. {Rafter}, R.~L. {Riddle}, S.~J. {Williams}, and D.~W.
  {Wingert}, \emph{\apj} \textbf{656}, 437--443 (2007)

\bibitem[{Casares} et~al.(2005)]{crp+05}
J.~{Casares}, I.~{Ribas}, J.~M. {Paredes}, J.~{Mart{\'{\i}}}, and C.~{Allende
  Prieto}, \emph{\mnras} \textbf{360}, 1105--1109 (2005)

\bibitem[{Dubus}(2006)]{dub06}
G.~{Dubus}, \emph{\aap} \textbf{456}, 801--817 (2006)

\end{thebibliography}

\IfFileExists{\jobname.bbl}{}
 {\typeout{}
  \typeout{******************************************}
  \typeout{** Please run "bibtex \jobname" to optain}
  \typeout{** the bibliography and then re-run LaTeX}
  \typeout{** twice to fix the references!}
  \typeout{******************************************}
  \typeout{}
 }

\end{document}